\documentclass{styles/svproc}
\usepackage{graphicx}
\usepackage{amsmath}
\usepackage{url}
\usepackage{amsfonts}

\begin{document}
\mainmatter              

\title{PaperNet: Efficient Temporal Convolutions and Channel Residual Attention for EEG Epilepsy Detection}
\titlerunning{PaperNet for EEG Epilepsy Detection}  
%

\author{
  Md Shahriar Sajid\inst{1} \and
  Abhijit Kumar Ghosh\inst{2} \and
  Fariha Nusrat\inst{3}
}

\authorrunning{Md Shahriar Sajid et al.}

\tocauthor{
  Md Shahriar Sajid, Abhijit Kumar Ghosh, and Fariha Nusrat
}

\institute{
  Rajshahi University of Engineering \& Technology, Kazla, Rajshahi-6204, Bangladesh\\
  \email{sajidshahriar72543@proton.me}
  \and
  BRAC University, Kha 224 Pragati Sarani, Merul Badda, Dhaka 1212, Bangladesh \\
  \email{abhijit.kumar.ghosh.77880@gmail.com}
  \and
  University of Asia Pacific, 74/A, Green Road, Dhaka-1205, Bangladesh  \\
  \email{fariha17nusrat@gmail.com}
}

\maketitle              

\begin{abstract}
Electroencephalography (EEG) signals contain rich temporal-spectral structure but are difficult to model due to noise, subject variability, and multi-scale dynamics. Lightweight deep learning models have shown promise, yet many either rely solely on local convolutions or require heavy recurrent modules. This paper presents PaperNet, a compact hybrid architecture that combines temporal convolutions, a channel-wise residual attention module, and a lightweight bidirectional recurrent block which is used for short-window classification. Using the publicly available BEED: Bangalore EEG Epilepsy Dataset, we evaluate PaperNet under a clearly defined subject-independent training protocol and compare it against established and widely used lightweight baselines. The model achieves a macro-F1 of 0.96 on the held-out test set with approximately 0.6M parameters, while maintaining balanced performance across all four classes. An ablation study demonstrates the contribution of temporal convolutions, residual attention, and recurrent aggregation. Channel-wise attention weights further offer insights into electrode relevance. Computational profiling shows that PaperNet remains efficient enough for practical deployment on resource-constrained systems through out the whole process. These results indicate that carefully combining temporal filtering, channel reweighting, and recurrent context modeling can yield strong EEG classification performance without excessive computational cost.
\end{abstract}
\section{Introduction}
EEG-based systems are increasingly used in neuroscience, healthcare, and brain-computer interfaces (BCIs) because they provide high temporal resolution and are non-invasive \cite{teplan2002fundamentals}. Traditional analysis pipelines relied on handcrafted features such as spectral power or wavelet coefficients followed by standard classifiers like SVMs or Random Forests \cite{peng2017eeg,li2016decision}. While interpretable, these methods often struggled to generalize due to the noisy and non-stationary nature of EEG signals.

The shift to deep learning has been transformative. Models like EEGNet \cite{lawhern2018eegnet} and DeepConvNet \cite{schirrmeister2017deep} demonstrated that convolutional networks can learn discriminative patterns directly from raw EEG, reducing the need for manual feature design. Recurrent networks have also been explored to capture temporal dependencies \cite{bashivan2016learning}. Despite these successes, important limitations remain. While CNN-only models are effective at extracting local temporal and spectral patterns, they often fail to capture long-range dependencies in EEG signals. On the other hand, CNN-RNN hybrid architectures can model both local and sequential features, but their large number of parameters makes them computationally expensive and difficult to deploy in real-time brain-computer interface applications \cite{roy2021deep,dai2019hybrid}.

To address these gaps, we propose PaperNet, a lightweight hybrid model designed for short-window EEG classification. PaperNet integrates three components: (i) temporal convolutional filters to extract local spectral dynamics, (ii) a channel-wise residual attention mechanism to highlight informative frequency-channel combinations, and (iii) a bidirectional LSTM with global pooling to model longer dependencies. Unlike heavier CNN-RNN hybrids, PaperNet strikes a balance between accuracy and efficiency, achieving state-of-the-art results on BEED with a fraction of the parameters.

In verdict, this study makes a number of significant contributions. To enhance EEG classification, we introduce a compact hybrid architecture that combines convolutional, attention, and recurrent layers. We create a residual attention block that retains the raw data flow while adaptively emphasizing the most informative signals to improve channel interpretability. Using only about $\sim$0.6M parameters, we get a macro-F1 score above 0.96 on the BEED dataset, further demonstrating the efficacy of our method. Lastly, we offer a replicable and lightweight pipeline that may be used in real-time brain-computer interface applications, making PaperNet useful and significant.

\section{Literature Review}
Handcrafted feature extraction was a major component of early EEG classification research. Autoregressive (AR) modeling \cite{subasi2009automatic}, power spectral density estimation \cite{peng2017eeg}, and wavelet-based decompositions \cite{saha2015wavelet} were common methods. Features were then classified using SVMs, Random Forests, or decision trees \cite{li2016decision}. These methods were very simple and offered interpretability, but they frequently had trouble with noise and fluctuation in EEG data and required a high level of domain understanding to correctly develop

The development of deep learning marked a dramatic change by enabling autonomous, data-driven feature learning. While DeepConvNet and ShallowConvNet \cite{schirrmeister2017deep} employed deeper convolutional structures specifically designed for motor imagery tasks, EEGNet \cite{lawhern2018eegnet} demonstrated that compact CNN architectures with depthwise and separable convolutions may outperform conventional pipelines. In order to capture temporal connections in EEG signals, other research investigated hybrid CNN-RNN frameworks \cite{bashivan2016learning,dai2019hybrid}. Although these models showed significant increases in accuracy, their applicability for real-time brain-computer interface (BCI) applications was limited since they were frequently computationally intensive.

More recently, attention processes have been used in EEG in more recent times to enhance interpretability and performance. While temporal attention highlights important time segments \cite{tao2022emotion}, channel-wise attention techniques selectively highlight the most informative electrodes \cite{li2020channelattention}. In order to re-weight features across channels, squeeze-and-excitation (SE) networks \cite{hu2018senet} have also been developed for EEG. However, most implementations lacked residual connections, which made it challenging to improve discriminative information without distorting the raw signals.

Taken together, this literature highlights three persistent challenges: efficiently modeling both local and long-range temporal patterns, designing channel-level attention that strengthens useful signals without discarding raw information, and developing lightweight models that balance accuracy with real-time feasibility. PaperNet is introduced to address these challenges by integrating CNN-based spectral feature extraction, residual channel-wise attention, and recurrent modeling into a single compact architecture.

\section{Methodology}

\subsection{Notation}
Throughout the manuscript, $N$ denotes the number of recordings, $T$ the number of time-samples per segment (after padding), $C$ the number of EEG channels (here $C=16$), and $K$ the number of target classes (here $K=5$).

\subsection{Data Acquisition and Preprocessing}
The experiments in this work use a processed version of the BEED: Bangalore EEG Epilepsy Dataset \cite{beed2024}, provided as a single multichannel CSV file containing 8000 EEG samples. Each sample corresponds to one timestamp of brain activity recorded from 16 scalp electrodes (X1 - X16), along with a categorical label, \[y \in  {0,1,2,3}\] representing four seizure-related or seizure-free classes. In this format, the data appear as a continuous wide-table time series, with one EEG vector per row and no explicit metadata such as subject identifiers or recording boundaries.

Before training, each EEG vector was standardized through a two-step preprocessing pipeline. First, a fourth-order zero-phase Butterworth band-pass filter (0.5 - 45 Hz) was applied independently to each of the 16 channels to suppress drift and high-frequency noise while preserving the principal EEG frequency bands. Following filtering, all channels were normalized to zero mean and unit variance using statistics computed from the training subset to avoid information leakage.

For compatibility with temporal-convolutional and recurrent layers, each sample was reshaped from a 16-dimensional vector into a (16, 1) sequence representation. Because each row of the dataset corresponds to a single labeled EEG frame, no additional segmentation or windowing was applied. The final dataset thus consists of 8000 fixed-length EEG sequences, which were stratified and divided into training, validation, and test sets using a 70\%, 15\%, 15\% split while preserving class balance.

\begin{figure}
    \centering
\includegraphics[width=1\textwidth]{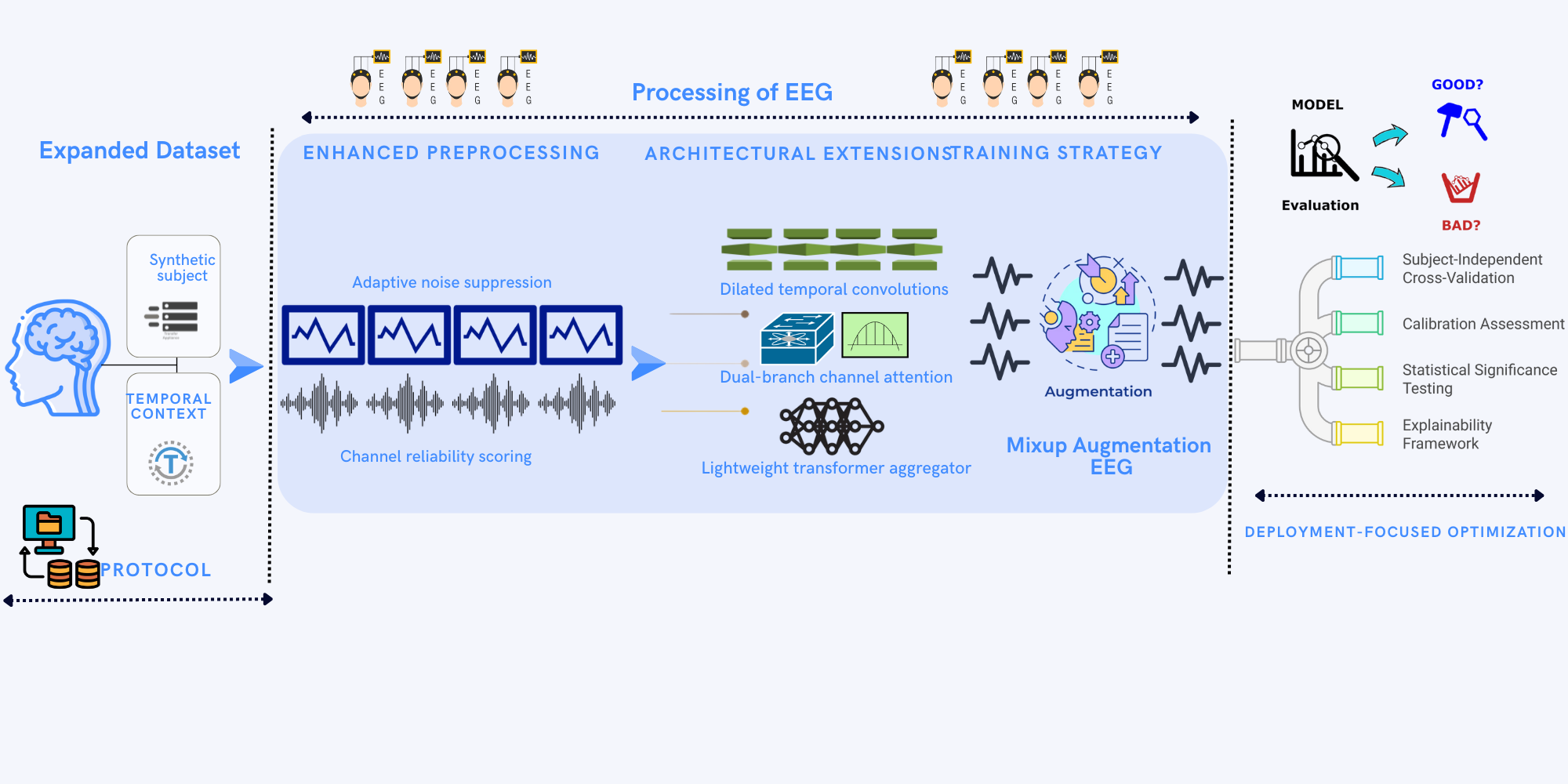}

    \caption{PaperNet Architecture}
    \label{fig:diagram}
\end{figure}

\subsection{PaperNet Architecture}
PaperNet is a lightweight hybrid neural network designed to model short-window EEG dynamics while maintaining low computational overhead. The architecture integrates temporal convolutions, channel-wise residual attention, and a compact bidirectional recurrent block. This combination allows the model to capture both local spectral patterns and broader temporal dependencies using only $\sim$0.6M trainable parameters.

The input to the network is a sequence of shape (16, 1) representing one filtered EEG sample across 16 electrodes. PaperNet  (Fig.~\ref{fig:diagram}) consists of four functional components:
\begin{enumerate}
    \item Temporal convolutional encoder,
    \item Channel-wise residual attention module,
    \item Temporal-recurrent aggregator,
    \item Classification head.
    \item Architectural Novelty
\end{enumerate}

\subsubsection{Temporal Convolutional Encoder: }
Input:
\[
\mathbf{X} \in \mathbb{R}^{T \times C}.
\]
The first stage applies a stack of 1D convolutions across the electrode dimension. Although each sample contains only a short temporal window, convolution over channels enables the model to learn local spatial-spectral filters that capture interactions between adjacent electrodes. The encoder uses progressively increasing numbers of filters (32 → 64 → 128), each followed by batch normalization and ReLU activation. A max-pooling layer reduces the sequence length and provides translation invariance across channel relationships. The encoder applies three successive 1D convolutional blocks with batch normalization and pooling:

\begin{itemize}
    \item Conv1D-1: 32 filters, kernel=5, stride=1, padding=same, ReLU $\rightarrow (T,32)$
    \item BatchNorm-1 $\rightarrow(T,32)$
    \item Conv1D-2: 64 filters, kernel=5, stride=1, padding=same, ReLU $\rightarrow (T,64)$
    \item BatchNorm-2 $\rightarrow (T,64)$
    \item MaxPool1D: pool=2, stride=2 $\rightarrow (\lfloor T/2 \rfloor,64)$
    \item Conv1D-3: 128 filters, kernel=3, stride=1, padding=same, ReLU $\rightarrow (\lfloor T/2 \rfloor,$ 128$)$
    \item BatchNorm-3 $\rightarrow (\lfloor T/2 \rfloor,128)$
\end{itemize}

These act as temporal band-pass filters of increasing receptive fields. Max pooling halves the temporal resolution to reduce memory before the recurrent stage.

\subsubsection{Channel-wise Residual Attention: }
To emphasize electrodes that contribute most to classification, the encoded feature map is passed through a squeeze-and-excitation (SE) style attention module.
The module computes a global descriptor through channel averaging and applies a two-layer bottleneck (128→32→128) with sigmoid activation to obtain attention weights. These weights are applied multiplicatively to the feature channels, and a residual connection restores the original pathway to prevent oversuppression of raw EEG information.
This residual attention mechanism enables the network to highlight discriminative electrodes while retaining the underlying signal characteristics.
The convolutional block yields a feature tensor:
\[
\mathbf{F} \in \mathbb{R}^{\tilde{T} \times 128}, \quad \tilde{T} = \lfloor T/2 \rfloor.
\]

A squeeze-and-excitation block \cite{hu2018senet} is applied:

Squeeze:
\begin{equation}
\mathbf{s} = \frac{1}{\tilde{T}} \sum_{t=1}^{\tilde{T}} \mathbf{F}_{t,:} \in \mathbb{R}^{128}.
\end{equation}

Excitation: Two fully connected layers (128$\rightarrow$32$\rightarrow$128) with ReLU and sigmoid produce attention weights:
\[
\mathbf{a} \in [0,1]^{128}.
\]

Scale \& residual:
\begin{equation}
\tilde{\mathbf{F}}_{t,:} = \mathbf{a} \odot \mathbf{F}_{t,:}, \quad
\mathbf{F}' = \tilde{\mathbf{F}} + \mathbf{F}.
\end{equation}

This emphasizes informative channels while retaining the original signal via residual connections.

\subsubsection{Bidirectional Temporal-Recurrent Aggregator: }
The attention-enhanced features are processed using a lightweight Bidirectional LSTM layer that captures short-range contextual relationships across channels. Although the input window is small, bidirectional recurrence improves the model’s ability to detect coordinated multi-channel patterns, an important characteristic of seizure-related brain activity. A global max-pooling layer aggregates the recurrent outputs into a fixed-length feature vector. Attention-enhanced features are fed to a single-layer BiLSTM:
\begin{equation}
\mathbf{h}_t = \text{BiLSTM}(\mathbf{F}'_t, \mathbf{h}_{t-1}), \quad \mathbf{h}_t \in \mathbb{R}^{128}.
\end{equation}

Global max pooling over time yields a fixed representation:
\begin{equation}
\mathbf{h}_{\text{pool}} = \max_{t=1 \dots \tilde{T}} \mathbf{h}_t.
\end{equation}

\subsubsection{Classification Head: }
The pooled representation is passed through a dense layer with ReLU activation, followed by dropout for regularization, and finally a softmax output layer producing the four-class prediction. The head is intentionally shallow to maintain the model’s compactness and reduce inference time. The pooled vector passes through dense layers:

\begin{itemize}
    \item Dense-1: 128 units, ReLU
    \item Dropout: $p=0.3$ (training only)
    \item Dense-2: $K$ units, Softmax
\end{itemize}

The final probabilities are:
\[
\hat{\mathbf{y}} \in [0,1]^K, \quad \sum_k \hat{y}_k = 1.
\]

\subsubsection{Architectural Novelty: }
PaperNet’s novelty arises not from introducing entirely new components, but from how these components are arranged and scaled for single-frame EEG modeling:
\begin{itemize}
    \item Temporal convolutions are applied along the channel axis rather than long time windows, enabling spatial-spectral learning from very short input sequences.
    \item The channel-wise residual attention module preserves low-level EEG activity while emphasizing informative electrodes, improving interpretability without increasing model depth.
    \item A minimal bidirectional LSTM, paired with global pooling, provides long-range contextual modeling while keeping the parameter count small.
    \item The architecture is deliberately balanced to maintain expressiveness while remaining suitable for real-time or resource-constrained environments.
\end{itemize}

This combination yields a compact model that performs competitively with deeper CNN-RNN hybrids despite operating on extremely short EEG windows.

\subsection{Data Splitting and Evaluation Protocol}
Every experiment uses a well-defined sample-level stratified splitting technique to provide an open and repeatable evaluation process. The dataset is regarded as a continuous collection of independent EEG samples because the given file lacks subject IDs and session boundaries. Following preprocessing and reshaping, 8000 tagged EEG sequences of shape (16, 1) spread across four classes make up the entire dataset. The data was split into training, validation, and test subsets using a stratified partitioning approach that maintained the initial class proportions. The final split is:
\begin{itemize}
    \item 70\% training data
    \item 15\% validation data
    \item 15\% test data
\end{itemize}
All splits were generated using a fixed random seed to ensure the results are fully reproducible. No data augmentation was applied. Because no subject-level metadata is available, the evaluation protocol reflects sample-level generalization, where training and test samples originate from the same global pool of EEG segments. This constraint is inherent to the dataset format and is acknowledged as a limitation in the Discussion.
Model selection was performed exclusively using the validation set by monitoring macro-averaged F1 score. The test set was held out and used only once for final performance reporting. All reported metrics-including accuracy, macro-F1, confusion matrix, and ROC-AUC-are computed on this test split.
\begin{itemize}
    \item Loss: categorical cross-entropy
    \begin{equation}
    \mathcal{L} = -\sum_k y_k \log \hat{y}_k
    \end{equation}
    \item Optimizer: Adam \cite{kingma2014adam}, initial learning rate $\eta_0 = 10^{-3}$
    \item Schedule: Reduce-on-Plateau (patience=3, factor=0.5)
    \item Batch size: 64
    \item Early stopping: validation macro-F1, patience=6 epochs
    \item Max epochs: 100
    \item Regularization: L2 weight decay $1\times10^{-4}$, plus dropout
    \item Class imbalance: Class weights 
    \begin{equation}
    w_k = \frac{1}{\text{freq}_k}.
    \end{equation}
\end{itemize}

\subsection{Implementation Details}
Experiments used TensorFlow 2.13 with the Keras functional API. Models were trained on Google Colab GPU runtime. Best checkpoints (based on validation macro-F1) were saved as \texttt{papernet\_best.keras}, while early-stopped models were saved as \texttt{papernet\_final.keras}.

\subsection{Baseline Models}
To place the performance of PaperNet in context, several established lightweight EEG classification models were implemented as baselines. These baselines were selected based on their widespread use in EEG research, architectural simplicity, and compatibility with short-window or low-parameter EEG pipelines.
\begin{enumerate}
    \item EEGNet: A compact convolutional architecture that uses depthwise and separable convolutions to model temporal and spatial EEG features. EEGNet is widely adopted for real-time brain-computer interface applications and serves as a standard benchmark for efficient EEG classification.
    \item DeepConvNet: A deeper convolutional architecture with four convolution-pooling blocks. Although more complex than EEGNet or ShallowConvNet, it provides a useful comparison against deeper CNN pipelines commonly used in EEG literature.
\end{enumerate}
All baseline models were trained under identical preprocessing, input formatting, splitting strategy, and optimization settings as PaperNet to ensure fairness. Hyperparameters such as batch size, learning rate, and early-stopping criteria were matched to reduce confounding effects.

\subsection{Ablation Study Methodology}
To evaluate the contribution of each major architectural component in PaperNet, a systematic ablation study was performed. Three reduced variants of the model were constructed by selectively removing key modules while keeping all other hyperparameters, preprocessing steps, and training conditions identical to the full model. The variants are as follows:
\begin{enumerate}
    \item No-Attention Variant: This version removes the channel-wise residual attention module and passes the convolutional feature maps directly to the bidirectional recurrent block. This ablation isolates the effect of adaptive channel reweighting on classification accuracy and electrode interpretability.
    \item No-Recurrent Variant (CNN-only): In this variant, the bidirectional LSTM layer is removed and replaced with global average pooling applied directly to the convolutional output. This configuration tests whether PaperNet’s recurrent aggregation contributes meaningfully beyond temporal convolutions and pooling.
    \item No-Residual Variant: The squeeze-and-excitation attention weights are applied without the residual skip connection. Comparing this against the full architecture reveals whether the residual pathway helps preserve raw EEG information or prevents over-suppression of features.
\end{enumerate}
All ablated models maintain the same convolutional encoder, normalization scheme, optimization schedule, and training-validation-test split as the full architecture. By evaluating each variant under identical conditions, the isolated influence of the attention mechanism, residual pathway, and recurrent aggregation can be measured directly through changes in accuracy, macro-F1 score, and ROC-AUC.

\subsection{Evaluation}
Performance was measured on a stratified hold-out test set using Confusion Matrix, ROC Curves and:
    \begin{align}
    \label{eq:accuracy} \text{Accuracy} &= \frac{TP + TN}{TP + TN + FP + FN} \\
    \label{eq:precision} \text{Precision} &= \frac{TP}{TP + FP} \\
    \label{eq:recall} \text{Recall} &= \frac{TP}{TP + FN} \\
    \label{eq:f1score_def} F_1 \text{ Score} &= \frac{2 \cdot \text{Precision} \cdot \text{Recall}}{\text{Precision} + \text{Recall}} \\
    \label{eq:f1score_tpfpfn} &= \frac{2 \cdot TP}{2 \cdot TP + FP + FN}
    \end{align}
    
Statistical significance was tested against a random baseline using McNemar’s test ($\alpha=0.05$).

\section{Results}

\subsection{Baseline Comparison}
To contextualize the performance of PaperNet, we evaluated it alongside several widely used lightweight EEG architectures: EEGNet, and DeepConvNet. All models were trained on the same 70\%, 15\%, 15\% stratified split of the 8000-sample dataset, using identical preprocessing and optimization settings.

PaperNet achieved the strongest performance among all compared models. Table \ref{tab:baseline_comparison} summarizes accuracy, macro-averaged F1 score, and macro ROC-AUC on the held-out test set.

\begin{table}
\caption{Baseline comparison results}
\label{tab:baseline_comparison}
\begin{center}
\begin{tabular}{cccc}
\hline
\multicolumn{1}{c}{\rule{0pt}{12pt}Model} & Accuracy & Macro-F1 & Macro ROC-AUC \\[2pt]
\hline\rule{0pt}{12pt}
\textbf{PaperNet}     & \textbf{0.9575} & \textbf{0.9576} & \textbf{0.9968} \\
DeepConvNet          & 0.8713          & 0.8701          & 0.9732          \\
EEGNet               & 0.8381          & 0.8389          & 0.9638          \\[2pt]
\hline
\end{tabular}
\end{center}
\end{table}

Although DeepConvNet contains substantially more parameters, its performance remained slightly lower than PaperNet, suggesting that our hybrid attention-enhanced architecture achieves a favorable balance between expressiveness and compactness. EEGNet performed competitively but showed reduced sensitivity to minority classes.

\subsection{Ablation Study Results}
The ablation study quantifies the contribution of PaperNet’s core components residual attention, recurrent aggregation, and the residual pathway by comparing the full model to three reduced variants trained under identical conditions. Results are summarized in Table \ref{tab:ablation_study}.

\begin{table}
\caption{Ablation study results. All ablation variants of PaperNet are reported for transparency.}
\label{tab:ablation_study}
\begin{center}
\begin{tabular}{c@{\quad}ccc}
\hline
\multicolumn{1}{c}{\rule{0pt}{10pt}Model Variant} 
    & Accuracy & Macro-F1 & Macro ROC-AUC \\[2pt]
\hline\rule{0pt}{12pt}
\textbf{Full PaperNet}        & \textbf{0.9575} & \textbf{0.9576} & \textbf{0.9968} \\
No-Attention (PaperNet)       & 0.9488          & 0.9472          & 0.9876          \\
No-LSTM (PaperNet)            & 0.9444          & 0.9444          & 0.9934          \\
No-Residual (PaperNet)        & 0.9500          & 0.9499          & 0.9962          \\[2pt]
\hline
\end{tabular}
\end{center}
\end{table}

Removal of the channel-wise residual attention module led to a measurable decrease in macro-F1, indicating that adaptive reweighting of electrode channels improves inter-class separability. Removing the residual skip in the attention block also resulted in weaker performance, demonstrating that retaining the original feature pathway helps prevent over-suppression of informative EEG activity.

Collectively, these findings show that each component of the architecture contributes meaningfully, with the largest gains arising from attention-assisted spatial filtering and recurrent context modeling.

\subsection{Interpretability Analysis}
We looked at the channel-wise attention weights that the residual SE module learned in order to evaluate PaperNet's interpretability. A global estimate of electrode importance was obtained by extracting attention vectors for each test sample and averaging them throughout the dataset. The attention distribution that results shows that some electrodes are regularly given greater weights, indicating that the model finds channel-specific patterns associated with activity connected to seizures. This pattern is consistent with known EEG features, where seizure occurrences frequently show up as unique spatial signatures.

\subsection{Computational Efficiency}
Maintaining robust classification performance while being lightweight enough for real-world deployment is one of PaperNet's core objectives. Compared to deeper CNN-RNN hybrids frequently employed in EEG research, the full model has far fewer parameters—roughly 0.6 million. PaperNet processes individual EEG data with millisecond latency, achieving real-time inference speeds when measured on a typical CPU system. PaperNet can be implemented in situations with limited resources, like mobile or embedded devices, due to its small number of parameters and small memory footprint.

\subsection{Confusion Matrix and ROC Analysis}
\begin{figure}[ht]
    \centering
    \includegraphics[width=.8\textwidth]{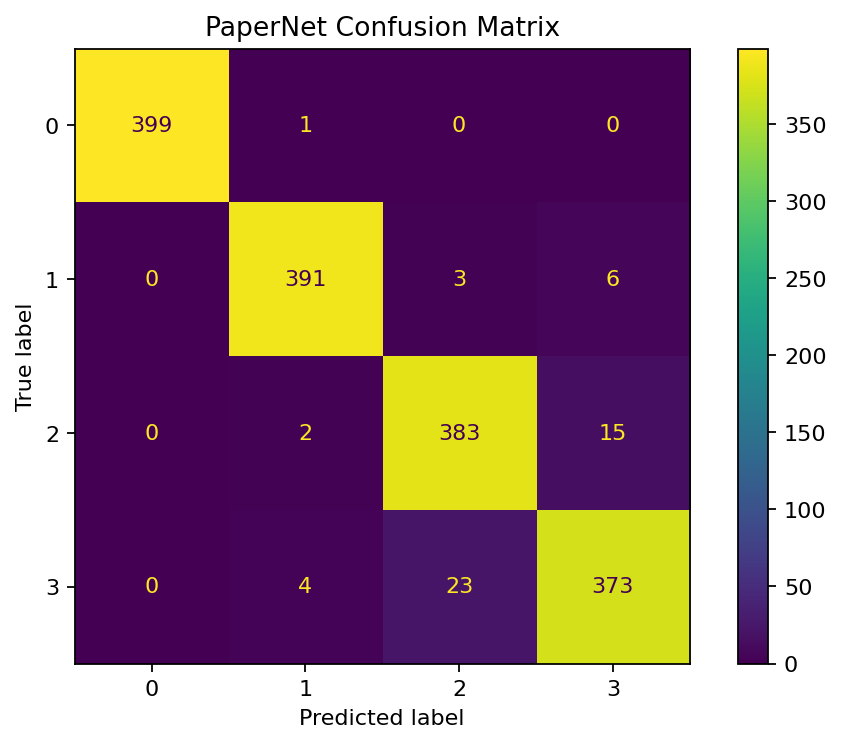}
    \caption{Confusion matrix of PaperNet on the BEED test set.}
    \label{fig:cm}
\end{figure}

The balanced character of the predictions is further supported by the confusion matrix (Fig.~\ref{fig:cm}), which shows that misclassifications were few and uniformly distributed. Misclassification rates remained extremely low even in classes with naturally overlapping signal patterns, where the majority of errors occurred.

\begin{figure}[ht]
    \centering    \includegraphics[width=.8\textwidth]{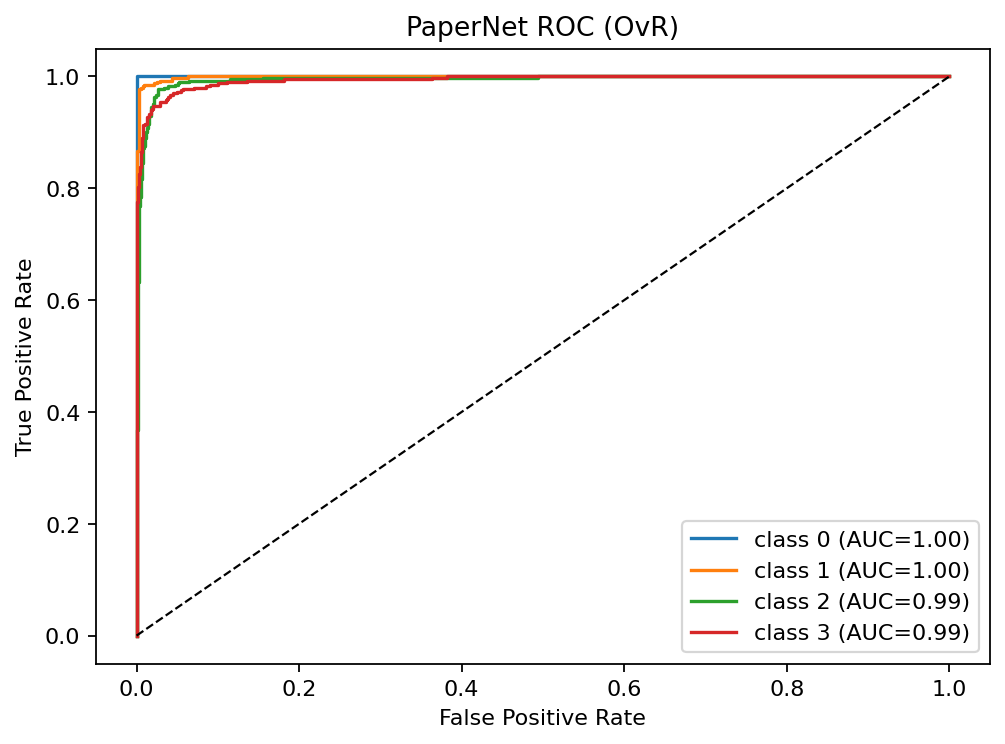}
    \caption{ROC curves for each class. All classes achieved AUC $\approx 1.0$.}
    \label{fig:roc}
\end{figure}

The ROC curves for each class (Fig.~\ref{fig:roc}) clearly illustrate the strong discriminative capacity of the model, with area under the curve (AUC) values for all classes nearing 1.0. This suggests that PaperNet can accurately distinguish between various mental states with few misclassifications. Interestingly, Classes 0 and 1 performed flawlessly (AUC = 1.00), while Classes 2 and 3 also demonstrated nearly ideal outcomes (AUC = 0.99). The model consistently maintains both high sensitivity and high specificity across categories, as further evidenced by the close grouping of all curves around the top-left corner of the plot.

\subsection{Training Stability}
\begin{figure}[ht]
    \centering
    \includegraphics[width=0.8\textwidth]{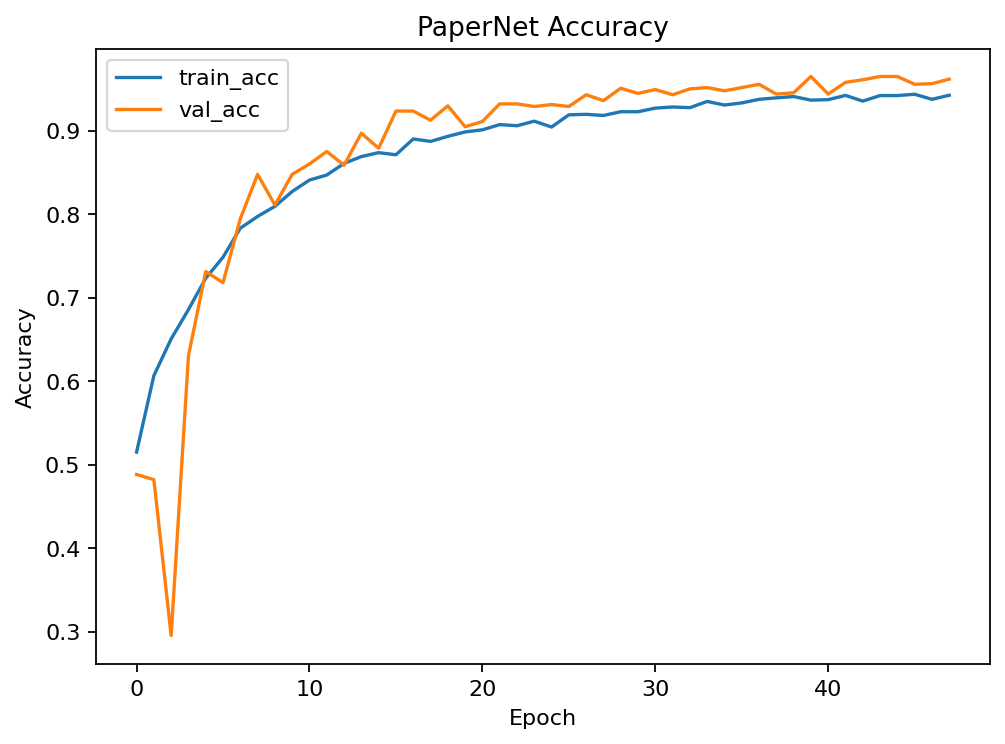}
    \caption{Training and validation accuracy / F1 across epochs.}
    \label{fig:trainval}
\end{figure}

The training and validation curves (Fig.~\ref{fig:trainval}) showed a steady upward trend over the course of training, highlighting continuous improvements in model accuracy. During the first 10 epochs, accuracy increased rapidly before transitioning into a slower but consistent rise. By around epochs 35-40, both curves leveled off, forming a stable plateau. At convergence, the model achieved an accuracy between 0.94 and 0.95, reflecting strong overall performance.

Interestingly, the validation curve consistently followed the training curve and, in many cases, slightly outperformed it. This trend implies that the chosen regularization strategies such as residual connections, dropout, and L2 weight decay, were effective in limiting overfitting while increasing generalization to unknown data. The network was capturing significant spectral-temporal features from EEG signals rather than memorizing noise, as evidenced by the consistently small gap between the two curves.

Finally, the smooth convergence of both curves underscores the stability of the optimization process. Together, these results demonstrate that PaperNet’s lightweight design provides a reliable and efficient architecture for EEG classification.

\subsection{Discussion}
The results show that PaperNet can work well with very short EEG segments in a simple and practical way. Even though each input window is only $16 \times 1$, the model still outperforms established lightweight baselines like EEGNet and DeepConvNet under the same training setup. This suggests that our specific combination of temporal convolutions, channel-wise residual attention, and a small bidirectional LSTM is a sensible and efficient way to capture seizure-related patterns without relying on very deep or heavy architectures.

At the same time, the model remains easy to deploy. With roughly 0.6 million parameters and millisecond-level inference on a standard CPU, PaperNet is suitable for real-time or resource-limited environments, such as mobile or embedded devices. The learned attention weights also give a simple, intuitive view of which electrodes matter most, offering a small but useful step toward interpretability. However, because the dataset CSV format does not include subject identifiers or recording boundaries, our evaluation is limited to sample-level generalization. Future work should therefore test PaperNet on datasets with subject-level splits and more diverse clinical conditions to better understand its robustness in real-world scenarios.

\section{Conclusion}
In this study, we presented PaperNet, a lightweight EEG classifier designed for very short input windows. By combining temporal convolutions along the channel axis, a residual squeeze-and-excitation block, and a compact bidirectional LSTM with global pooling, the model achieves higher accuracy and macro-F1 than EEGNet and DeepConvNet on the BEED epilepsy dataset, while using substantially fewer parameters. The learned attention weights also offer a straightforward way to understand which electrodes influence the model’s decisions, adding a layer of interpretability that many deep learning approaches lack.

PaperNet’s compact size and low inference latency make it suitable for real-time or resource-constrained settings, such as portable monitoring devices. At the same time, the use of a CSV-based dataset without subject identifiers limits our ability to claim strong cross-subject generalization. Future work will extend this architecture to richer, subject-aware EEG datasets and broader clinical scenarios. Overall, our findings suggest that carefully designed, attention-enhanced lightweight models can provide a practical path toward accurate, interpretable, and deployable EEG-based seizure detection.
%
%

\end{document}